\documentclass[pra,balancelastpage,twocolumn,footinbib,superscriptaddress,showpacs]{revtex4-1}
\usepackage{color,amsthm,amsmath,amsfonts,graphicx,bm}
\usepackage{bm}
\usepackage{color}
\usepackage{amsfonts}
\usepackage{amssymb}
\usepackage{amsmath}
\usepackage{epsfig}
\usepackage{supertabular}
\usepackage{graphicx}
\usepackage{enumerate}
\usepackage{multirow}
\usepackage{verbatim}
\usepackage{color}
\usepackage{subfigure}
\usepackage[T1]{fontenc}
\usepackage{array}
\usepackage{booktabs}
\usepackage{tabularx}
\begin{document}
\title{Exploring Magnetic Phases in Dual-Species Mott insulating Spinor Lattice Gases}

\author{Rui-Shan Li}
\affiliation{College of Physics and Electronic Engineering, Sichuan Normal University, Chengdu 610068, China}
\author{Zong-Zhen Pan}
\affiliation{Center for Correlated Matter and School of Physics, Zhejiang University, Hangzhou 310058, China}
\author{Shi-Jie Yang}
\affiliation{Department of Physics, Beijing Normal University, Beijing 100875, China}
\author{Yi Zheng}
\email{Corresponding author: zhengyireal@sicnu.edu.cn}
\affiliation{College of Physics and Electronic Engineering, Sichuan Normal University, Chengdu 610068, China}


\begin{abstract}
We explore the Mott insulating phases of dual-species bosonic spinor lattice gases, emphasizing the intriguing interplay between synthetic flux and inter-species spin exchange interaction. One of the species is subjected to Raman assisted tunneling, which leads to a synthetic flux within the framework of synthetic dimensions. In the deep Mott regime, the low energy physics is governed by an unconventional and highly tunable spin model, which is characterized by two distinct spin chains. The synthetic flux serves as an effective spin-orbit coupling, inducing Dzyaloshinskii-Moriya interactions in one of the spin chains. The inter-species spin exchange interaction gives rise to the inter-chain coupling embodied as an isotropic XX interaction. Using time-evolving block decimation method for tensor network states, we compute order parameters, correlation functions and structure factors to identify the ground state magnetic phases. The DM interaction in one species, when combined with the inter-species spin-exchange interaction, can induce spiral magnetic order in the second, otherwise non-chiral species. Besides, the interplay of a transverse field applied to one spin chain and the inter-species coupling can drive both spin chains into a paramagnetic phase simultaneously. These results reveal that inter-species coupling serves as a powerful conduit for transmitting magnetic correlations, enabling exotic phases beyond the single-component perspective.
\end{abstract}

\maketitle

\section{Introduction}\label{intro}
Ultracold atomic system has provided a versatile platform of quantum simulation, allowing for the study of various quantum many-body phenomena ranging from superfluidity to quantum magnetism in the Mott insulating phase \cite{jaksch2005cold, lewenstein2007ultracold, bloch2008many, gross2017quantum, schafer2020tools, greiner2002quantum, jordens2008mott, simon2011quantum, hilker2017revealing}. By loading spinor quantum gases in one dimensional (1D) optical lattice, recent experiments have implemented artificial gauge fields via Raman-assisted tunneling \cite{celi2014synthetic, mancini2015observation, stuhl2015visualizing}. The pseudo-spin degree of freedom is mapped to a synthetic space dimension, converting the 1D system to a ladder with synthetic flux. The spin-flip process is interpreted as the rung coupling. Such a system serves as a fundamental testbed for exploring Hofstadter fractal spectrum \cite{hofstadter1976energy, dalibard2011colloquium, aidelsburger2013realization, miyake2013realizing}, Hall edge current \cite{mancini2015observation, stuhl2015visualizing, greschner2015spontaneous, wei2014theory, cornfeld2015chiral}, Meissner-vortex phase transition \cite{atala2014observation, di2015meissner, piraud2015vortex, zhou2023spin, impertro2025strongly}, and effective spin-orbit coupling (SOC) \cite{wang2010spin, galitski2013spin, zhai2015degenerate, wu2016realization} in neutral quantum gases . In the Mott regime, charge fluctuations are suppressed and spin exchange dominates the low-energy physics. Within second order perturbation, the system can be described by an effective spin Hamiltonian such as the anisotropic Heisenberg model \cite{kuklov2003counterflow, duan2003controlling}. The implementation of artificial gauge fields further induces Dzyaloshinskii-Moriya (DM) interaction \cite{dzyaloshinsky1958thermodynamic, moriya1960}, which is well-established in strongly correlated electronic materials and plays a crucial role in stabilizing chiral magnetic textures like skyrmions and spin spirals \cite{coffey1991dzyaloshinskii, di2015direct, yang2023first, cole2012bose, radic2012exotic, cai2012magnetic, xu2014mott, gong2015dzyaloshinskii, fert2017magnetic}. By adjusting lattice parameters and the synthetic flux, various spin models can be covered with highly tunable spin-spin couplings, external fields, and the DM interaction. The effects of DM interaction and anisotropy may lead to diverse magnetic phases beyond the conventional ferromagnetic and antiferromagnetic orders \cite{zhang2019magnetic, sun2021effective, pan2024effective}. Typical examples include the gapless spiral phase in 1D spin chains \cite{xu2014mott} and vortex crystals (or even Skyrmion) in two dimensional systems \cite{cole2012bose, radic2012exotic}. 

On the other hand, a system comprising two distinct atomic species offers a rich playground for exploring interacting quantum matter, multicomponent superfluidity and exotic phases. Dual-species mixtures have been realized in cold atom experiments, either in a magneto-optical trap \cite{hall1998dynamics, maddaloni2000collective, ospelkaus2010quantum, ferrier2014mixture, yao2016observation, roy2017two} or in an optical lattice \cite{catani2008degenerate,roati2008anderson}. Dynamical effects driven by interspecies spin-spin interactions have also been reported \cite{li2015coherent}. Interestingly, for a system of dual-species spinor superfluid mixture, it has been shown that an effect of SOC can be transmitted from one component to the other through inter-species spin-exchange interaction \cite{chen2018spin, zhu2019spin}. These studies have revealed an extensive landscape of phases and phenomena that are absent in single-species systems. 

However, most previous studies concerning the mixture of quantum gases are focused on the superfluid phase, the features of spin ordering in the deep Mott regime have attracted less attention. In the present work, we consider the mixture of spinor quantum gases with strong interactions loaded in a 1D optical lattice. Each species constitutes two hyperfine states, forming a ladder-like lattice in the spirit of synthetic dimension \cite{celi2014synthetic, ozawa2019topological}. Only one of the species is directly subjected to Raman lasers, resulting in a synthetic flux, which further induces DM interaction in the strong interaction limit. At this point, the system can be mapped to a two-layer spin model. With the inter-layer coupling caused by the inter-species spin-exchange interaction, we expect that the spiral order can be induced in both layers, leading to intriguing phases that are distinct from the conventional case of coupled spin chains.

The article is organized as follows. In Sec. \ref{II}, we present the dual-species spinor lattice model within the framework of Bose-Hubbard Hamiltonian. In the deep Mott insulating regime, such a model is mapped to an effective spin Hamiltonian under second-order perturbation, resulting in a system of two coupled spin chains. In Sec. \ref{III}, we study the ground state phases and phase transitions under certain conditions. We observe spiral and paramagnetic phases that are induced by the inter-species spin exchange interaction. A summary is included in Sec. \ref{IV}

\section{Two-species ladder model and effective spin hamiltonian}\label{II}

\begin{figure}[tbb!]
	\vspace{1em}
	\centering
	\includegraphics[width=.4\textwidth]{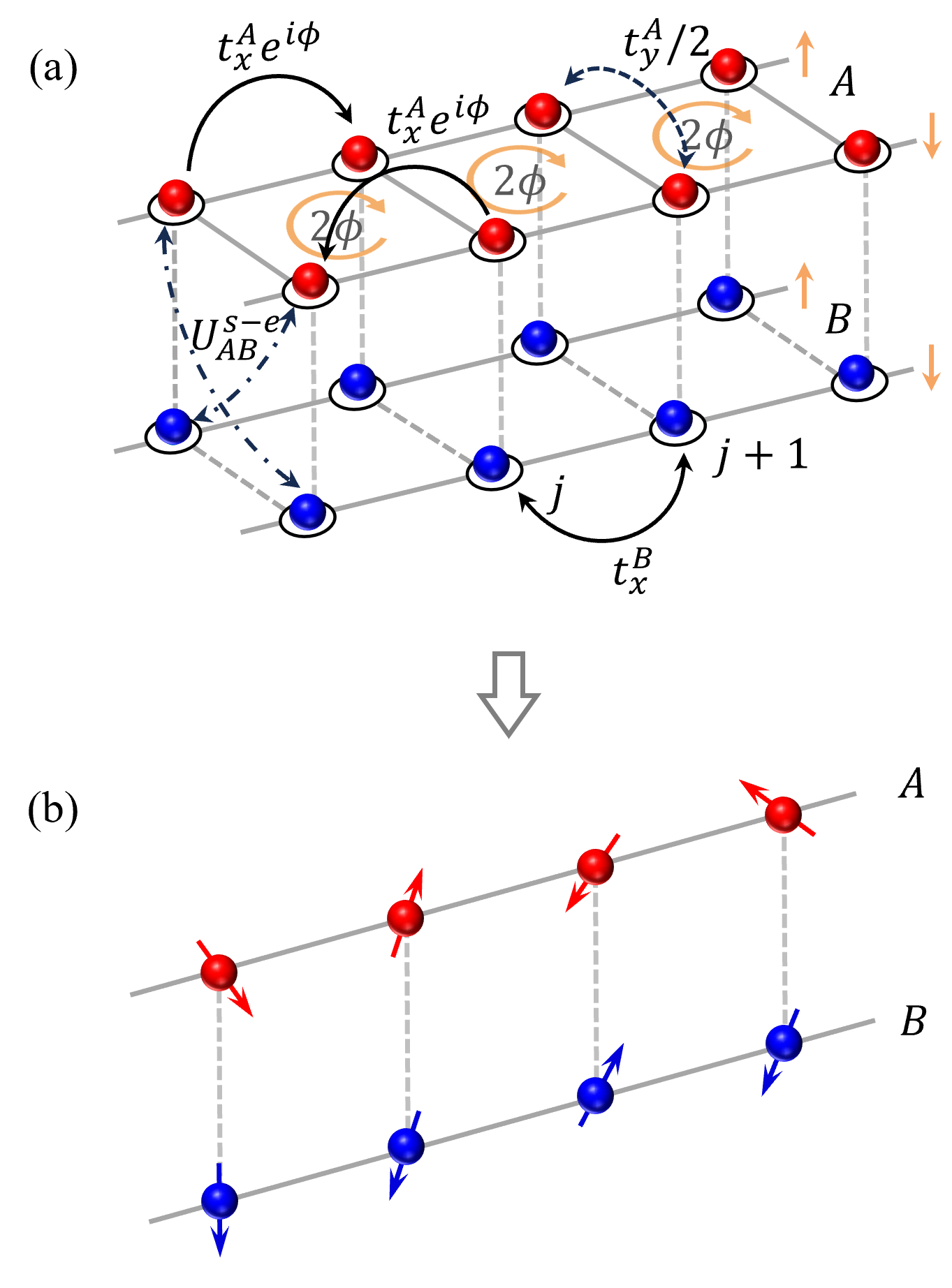}
	\caption{(a) Schematic diagram of the two-layer bosonic ladder system described by Eq. (\ref{eq_H}). Pseudospins are labeled by $\uparrow$ and $\downarrow$, corresponding to the ladder legs. Species A is subjected to a Raman assisted tunneling, constructing a chiral ladder with synthetic flux $2\phi$ per plaquette. In contrast, species B is not coupled to the Raman field, resulting in a flux-free ladder. (b) Schematic illustration of the effective spin model (\ref{eq_spin}), which represents a system of two coupled spin chains.
	\label{fig1}
	}
\end{figure}

We consider a mixture of two-species bosonic quantum gases (labeled as $A$ and $B$) loaded in a 1D optical lattice. Each species consists of two internal states labeled by $\uparrow$ and $\downarrow$, as represented by the two legs in Fig. 1(a). The link between $\uparrow$ and $\downarrow$ represent spin-flip processes. Species $A$ is engineered by Raman-assisted tunneling, leading to spin-dependent hoppings, whereas species $B$ is not directly coupled to the Raman lasers. Therefore, the upper layer in Fig. 1(a) constitutes a two-leg ladder with a synthetic flux $2\phi$ and the lower layer is a plain ladder. The total Hamiltonian is written as
\begin{equation}\label{eq_H}
\hat H = \sum_{i=A,B} \left(\hat H_i + \hat H_U^i \right) + \hat H_{AB}.
\end{equation}
The single-particle tunneling terms for species $A$ and $B$ are respectively given by
\begin{eqnarray}\label{eq_Ha}
\hat H_A =& -&t_x^A \sum_j \left( e^{i\phi} \hat a_{j+1,\uparrow}^{\dagger} \hat a_{j,\uparrow} + e^{-i\phi} \hat a_{j+1,\downarrow}^{\dagger} \hat a_{j,\downarrow}\right) \nonumber\\
&-& \frac{t_y^A}{2} \sum_j  \hat a_{j+1,\uparrow}^{\dagger} \hat a_{j,\downarrow} + \text{H.c.},
\end{eqnarray}
\begin{equation}\label{eq_Hb}
\hat H_B = - t_x^B \sum_j \left( \hat b_{j+1,\uparrow}^{\dagger} \hat b_{j,\uparrow} + \hat b_{j+1,\downarrow}^{\dagger} \hat b_{j,\downarrow}\right) + \text{H.c.}.
\end{equation}
Here $a_{j,\sigma }^{\dag }$ and $b_{j,\sigma }^{\dag }$ are creation operators for a particle of species $A$ and $B$ with spin $\sigma = \uparrow, \downarrow$. $t_x^i$ ($i=A,B$) denotes the tunneling amplitude along the lattice. $t_y^A$ controls the spin-flip hopping rate. For species A, the hopping phase $\phi$ produces a net $2\phi$ flux per plaquette, as implemented in recent cold atom experiments \cite{atala2014observation, mancini2015observation, stuhl2015visualizing}. Such a chiral ladder is featured by an effective SOC \cite{hugel2014chiral, junemann2017exploring}, and has attracted much attention in the study of superfluid vortex \cite{atala2014observation, di2015meissner}, as well as magnetic ordering in the insulating phase \cite{zhang2019magnetic, sun2021effective, pan2024effective}. 

The interaction for species $i$ in Eq. (\ref{eq_H}) is written as
\begin{equation}
\hat H_U^i = U_{\uparrow\downarrow}^i \sum_j \hat{n}_{j,\uparrow}^i \hat{n}_{j,\downarrow}^i + \frac{1}{2} \sum_{j,\sigma} U_\sigma^i \hat{n}_{j,\sigma}^i (\hat{n}_{j,\sigma}^i - 1),
\end{equation}
where $U_{\uparrow\downarrow}^i$ and $U_\sigma^i$ denote interspin and intraspin interaction strengths, respectively. The operator $\hat{n}_{j,\sigma}^i$ gives the particle number at site $j$ for spin-$\sigma$ atoms of species $i$. The interaction between $A$ and $B$ is governed by
\begin{eqnarray}
\hat H_{AB} &= &U_{AB}^\text{d-d}\sum_j \hat n_j^A\hat n_j^B \nonumber\\
&+&U_{AB}^\text{s-e} \sum_j \left( \hat a_{j,\uparrow}^{A\dagger} \hat a_{j,\downarrow}^A \hat a_{j,\uparrow}^{B\dagger} \hat a_{j,\downarrow}^B + \text{H.c.} \right),
\end{eqnarray}
which is the result of second-order quantization by taking single band approximation for localized Wannier states \cite{jaksch1998cold, zhai2021ultracold}. Here $U_{AB}^\text{d-d}$ characterizes the spin-independent density-density interaction with $\hat n_j^A=(\hat a_{j,\uparrow}^\dag \hat a_{j,\uparrow}+\hat a_{j,\downarrow}^\dag \hat a_{j,\downarrow})$ and $\hat n_j^B=(\hat b_{j,\uparrow}^\dag \hat b_{j,\uparrow}+\hat b_{j,\downarrow}^\dag \hat b_{j,\downarrow})$. The second term represents the inter-species spin-exchange interaction. To simplify the model, we assume that the inter-species interactions are spin-independent, with $U_\uparrow^i=U_\downarrow^i \equiv U_i$. Furthermore, we set all interactions positive and $U_{AB}^\text{d-d}<\sqrt{U_AU_B}$, in order to make the system stable and miscible. 

We focus on the deep Mott regime at unit filling ($n_A = n_B = 1$) for both species. With $U_i,U_{\uparrow\downarrow}^i\gg t_x^i, t_y^i, U_{AB}^\text{s-e}$, the low-energy physics is dominated by virtual tunneling and spin-exchange processes. The unperturbed Hamiltonian $H_0 = \sum_i H_U^i$ possesses a highly degenerate ground-state manifold with single occupation of both $A$ and $B$ particles per site. The perturbation term $H' = \sum_i H_i + H_{AB}$ couples these states through virtual excitations. Up to second order, the effective Hamiltonian can be obtained by:
\begin{equation}
\langle p|\hat H_{\rm eff}|p'\rangle = E_p^0\delta_{pp'} + \frac{1}{2}\sum_d \frac{\langle p|\hat H'|d\rangle\langle d|\hat H'|p'\rangle}{E_p^0 - E_d^0},
\end{equation}
where $|p\rangle$ and $|p'\rangle$ denote two degenerate states in the ground-state manifold. $|d\rangle$ represents an excited state of double-occupancy. $E_p^0$ and $E_d^0$ correspond to the eigenvalues of $|p\rangle$ and $|d\rangle$ respectively. An effective spin Hamiltonian emerges by defining the spin operators as $\mathbf{S}_j^i = \frac{1}{2} a_{j,\alpha}^{i\dagger} \boldsymbol{\sigma}_{\alpha\beta} a_{j,\beta}^i$, where $\boldsymbol{\sigma} = (\sigma^x, \sigma^y, \sigma^z)$ denotes the Pauli matrices. The effective Hamiltonian takes the form
\begin{eqnarray}\label{eq_spin}
\hat H_\text{eff} &=& \sum_{k = A,B}\sum_{j}\sum_{a=x,y,z} {J_{a,k}} \hat S_j^{a,k} \hat S_{j + 1}^{a,k} \nonumber\\
&+& \sum\limits_j \left[ \vec{D}_A \cdot (\vec{S}_j^A \times \vec{S}_{j + 1}^A) \right. \nonumber\\
&-& \left. h_{x,A} \hat S_j^{x,A} + J_{AB} \left( \hat S_j^{x,A} \hat S_j^{x,B} + \hat S_j^{y,A} \hat S_j^{y,B} \right) \right].
\end{eqnarray}
Here we have neglected a trivial energy shift $\Delta E=U_{AB}^\text{d-d}$. Such a Hamiltonian describes a dual spin chain model as illustrated in Fig. 1(b). The first line in Eq.(\ref{eq_spin}) represents spin-spin coupling. $\vec D_A$ and $h_x$ terms respectively refer to the DM interaction and a transverse field acting only on the $A$-chain. As indicated by the $J_{AB}$ term, the two chains are connected through an isotropic XX interaction, which stems from the inter-species spin exchange. For convenience, we scale the energy in units of $4(t_x^A)^2/U_A$ and set $U_{\uparrow\downarrow}^i = \lambda_i U_i$. We further define a dimensionless ratio $\beta \equiv {(t_x^B)^2 U_A}/{(t_x^A)^2 U_B}$ which quantifies relative energy scales between the two species. The spin-spin coupling strengths $J_{a,k}$  are given by
\begin{eqnarray}
&J_{x(y),A} &=  - \frac{1}{\lambda_A}\cos 2\phi, \nonumber\\
&J_{z,A} &=  - \frac{1}{\lambda_A}(2\lambda_A - 1),\nonumber\\
&J_{x(y),B} &=  - \frac{\beta }{\lambda _B},\nonumber\\
&J_{z,B} &=  - \frac{\beta }{\lambda _B}(2\lambda_B - 1),
\end{eqnarray}
indicating an interaction of the anisotropic Heisenberg type for both spin chains. Other non-zero couplings in Eq.(\ref{eq_spin}) are
\begin{eqnarray}
&D_{z,A}& =  - \frac{1}{\lambda _A}\sin 2\phi \nonumber\\
&h_{x,A}& =   {t_y^A}/(\frac{{4(t_x^A)}^2}{U_A})\nonumber\\
&J_{AB}& =   2U_{AB}^\text{s-e}/(\frac{{4(t_x^A)}^2}{U_A}).
\end{eqnarray}
In general, the two chains are governed by XXZ interaction. Taking $\phi$ and $\lambda_i$ as tuning parameters, the Ising-type interaction, the XX model, as well as the isotropic Heisenberg model can also be covered. The only appearance of $z$-component in DM interaction is guaranteed by the symmetry of the original Hamiltonian. Specifically, we take $M$, $\sigma_x$ and $K$ as mirror reflection ($M: j\leftrightarrow -j$), spin reversal ($\sigma_x: \uparrow\leftrightarrow\downarrow$) and complex conjugation, respectively. The Hubbard model is then invariant under $M\sigma_x$, $MK$ and $K\sigma_x$ acting on both species, whereas the $x$- and $y$-component of the DM interaction are anti-symmetric under these operations.

\section{magnetic phases tuned by spin exchange interaction}\label{III}
To characterize the ground state magnetic ordering and phases, we focus on the following quantities: (i) spin-spin correlation function 
\begin{equation}
\mathcal S_{j,l}^{a,k} = \left\langle \hat S_j^{a,k} \hat S_l^{a,k} \right\rangle,
\end{equation}
and the associated structure factor
\begin{equation}
S_a^k(q) = \frac{1}{L}  \sum_{j,l} e^{iq(j-l)} \mathcal S_{j,l}^{a,k},
\end{equation}
where $a \in \{x, y, z\}$, $k \in \{A,B\}$, $L$ is the lattice size and $q$ is the wave vector; (ii) long-range ferromagnetic correlations 
\begin{equation}
M_a^k = \frac{1}{2d+1} \sum_{r} \left\langle \hat S_j^{a,k} \hat S_{j+r}^{a,k} \right\rangle_j
\end{equation}
and long-range antiferromagnetic correlations
\begin{equation}
N_a^k = \frac{1}{2d+1} \sum_{r} (-1)^r \left\langle \hat S_j^{a,k} \hat S_{j+r}^{a,k} \right\rangle_j,
\end{equation}
which can be taken as order parameters for ferromagnetic phase and antiferromagnetic phase; (iii) long-range spiral correlations
\begin{equation}
C_a^k = \frac{1}{2d+1} \sum_{r} \left\langle \left[\vec{S}_j^k \times \vec{S}_{j+r}^k\right]^a \right\rangle_j,
\end{equation}
which capture the character of spiral ordering.
Here $\langle \cdot \rangle_j$ denotes averaging over lattice site $j$. The summation for $r$ is taken over the interval $[r_0-d,r_0+d]$ with $r_0=L/2$ and $d = L/4$. We mention that $M_a^k$, $N_a^k$ and $C_a^k$, which capture the features of ferromagnetic, antiferromagnetic and spiral ordering, are defined separately for the two spin chains.

To numerically calculate the expectation value of these quantities, we take time-evolving block decimation (TEBD) method based on tensor network (TN) representations \cite{schollwock2011density, orus2014practical, orus2014advances, hauschild2018efficient, banuls2023tensor}, which is recognized for its versatility, high precision and computational efficiency. Ground states for Hamiltonian (\ref{eq_spin}) are calculated on an open-boundary lattice with lattice size $L=160$. The virtual bond dimension of the TN is constrained to D=60.

Without the $J_{AB}$ term in Eq. (\ref{eq_spin}), the $A$-chain is an XXZ model with a $z$-component DM interaction and a transverse field. The ground state exhibits a variety of magnetic phases and may show long-range spiral correlations. For $\phi=\pi/4$ and $h_{x,A}= 0$, such a spin model can be addressed by quantum renormalization group approach \cite{jafari2008phase}. The ground state experiences a phase transition between a ferromagnetic phase along $z$-direction and a spiral phase with algebraically decaying spiral correlations. The phase boundary is located at $\lambda_A=1$. Under a weak transverse field, the spiral phase is then featured by long-range correlations and the boundary slightly deviates from $\lambda_A=1$. 

\begin{figure}[tbp]
	\centering
	\includegraphics[width=.49\textwidth]{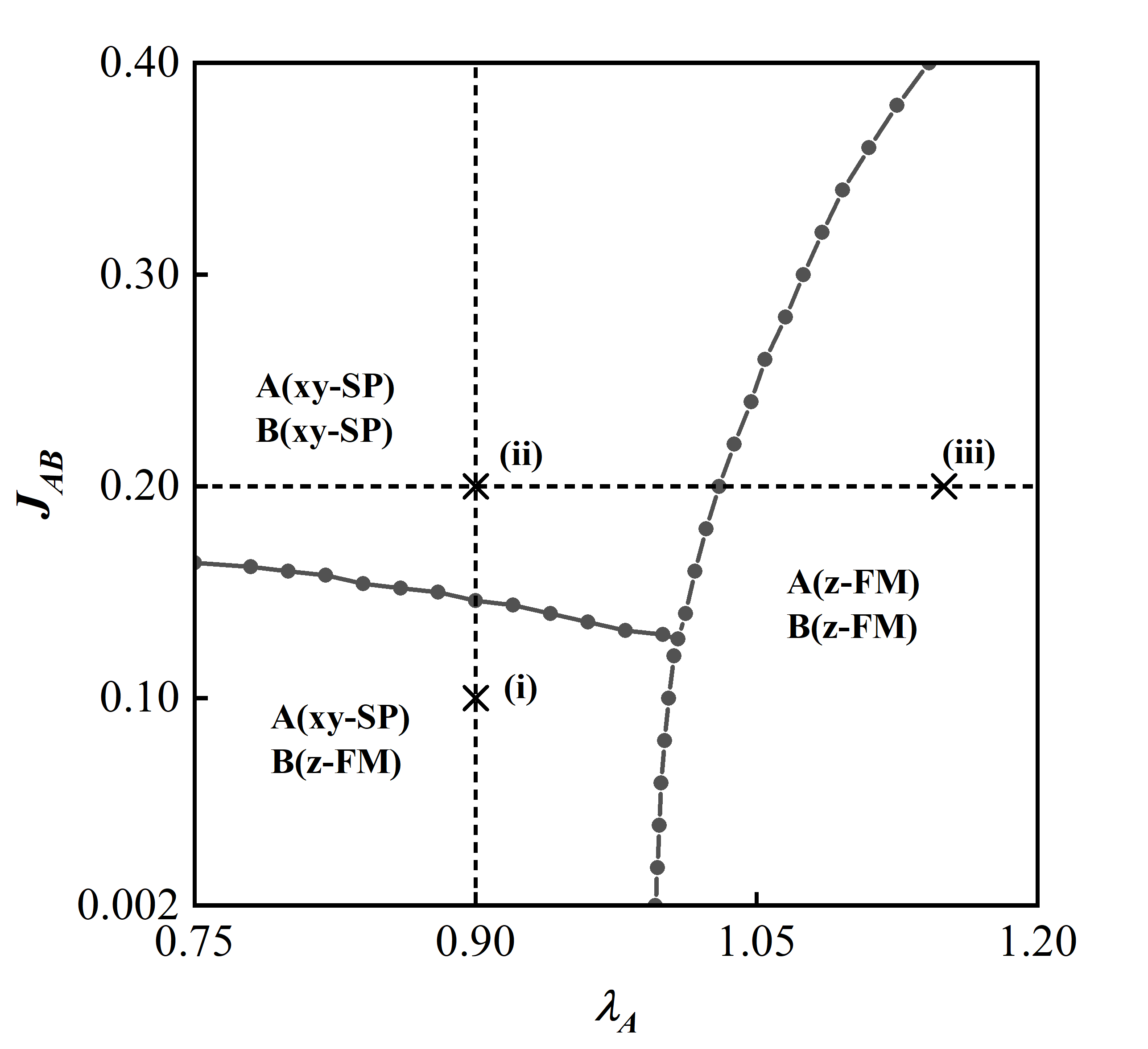}
	\caption{Ground state phase diagram of the spin model governed by Hamiltonian (\ref{eq_spin}) with dimensionless parameters $(\phi, h_{x,A}, \lambda_B, \beta) = ( 0.25\pi, 0.1, 1.2,0.1)$. Three phases are labeled in the diagram. The increase of the inter-chain coupling $J_{AB}$ induces the transition to $xy$-SP phase for species $B$. Phase transitions along the dashed lines are demonstrated in Fig. 3. Typical points marked with crosses are analyzed with correlation functions and structure factors in Fig. 4.
	}
	\label{fig.2}
\end{figure}

\begin{figure}[tbp]
	\includegraphics[width=.5\textwidth]{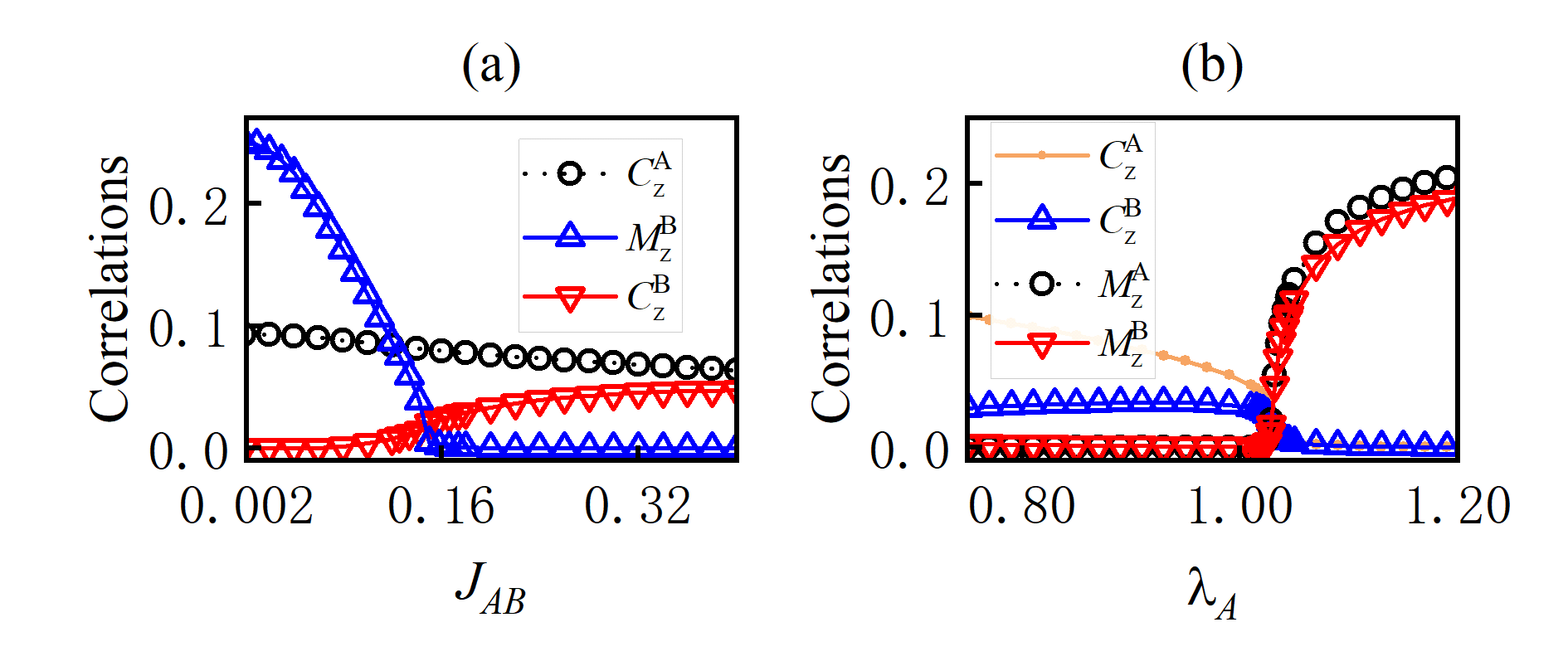}
	\caption{Long-range correlations as a function of $J_{AB}$ and $h_{x,A}$ along the dashed lines in fig. 2. (a) $C_z^A$, $M_x^B$ and $C_z^B$ versus $J_{AB}$ with $\lambda_A =0.9$. (b) $C_z^{A(B)}$ and $M_z^{A(B)}$ versus $\lambda_A$ with $J_{AB}=0.2$.
	}
	\label{fig.3}
\end{figure}

\subsection{spiral phase induced by inter-species spin-exchange interaction with a weak field}
We investigate the influence of interspecies spin-exchange interaction $U_{AB}^\text{s-e}$, namely the inter-chain coupling $J_{AB}$, on the $B$-chain and magnetic phases. We set $(\phi,h_{x,A})=(0.25\pi,0.1)$ and $\lambda_A\in [0.75,1.2]$ so that the phase transition between the ferromagnetic phase and the spiral phase is covered \cite{zhang2019magnetic}. For the $B$-chain, we take the XXZ Heisenberg interaction with $\lambda_B=1.2$. $\beta = 0.1$ is selected to ensure the sensitivity to the influence of the $A$-chain. As the inter-chain coupling energetically favors antiparallel alignment of the two spin chains in the $x$-$y$ plane, an induced spiral order in species $B$ can be expected. 

\begin{figure*}[tbp]
	\includegraphics[width=0.98\textwidth]{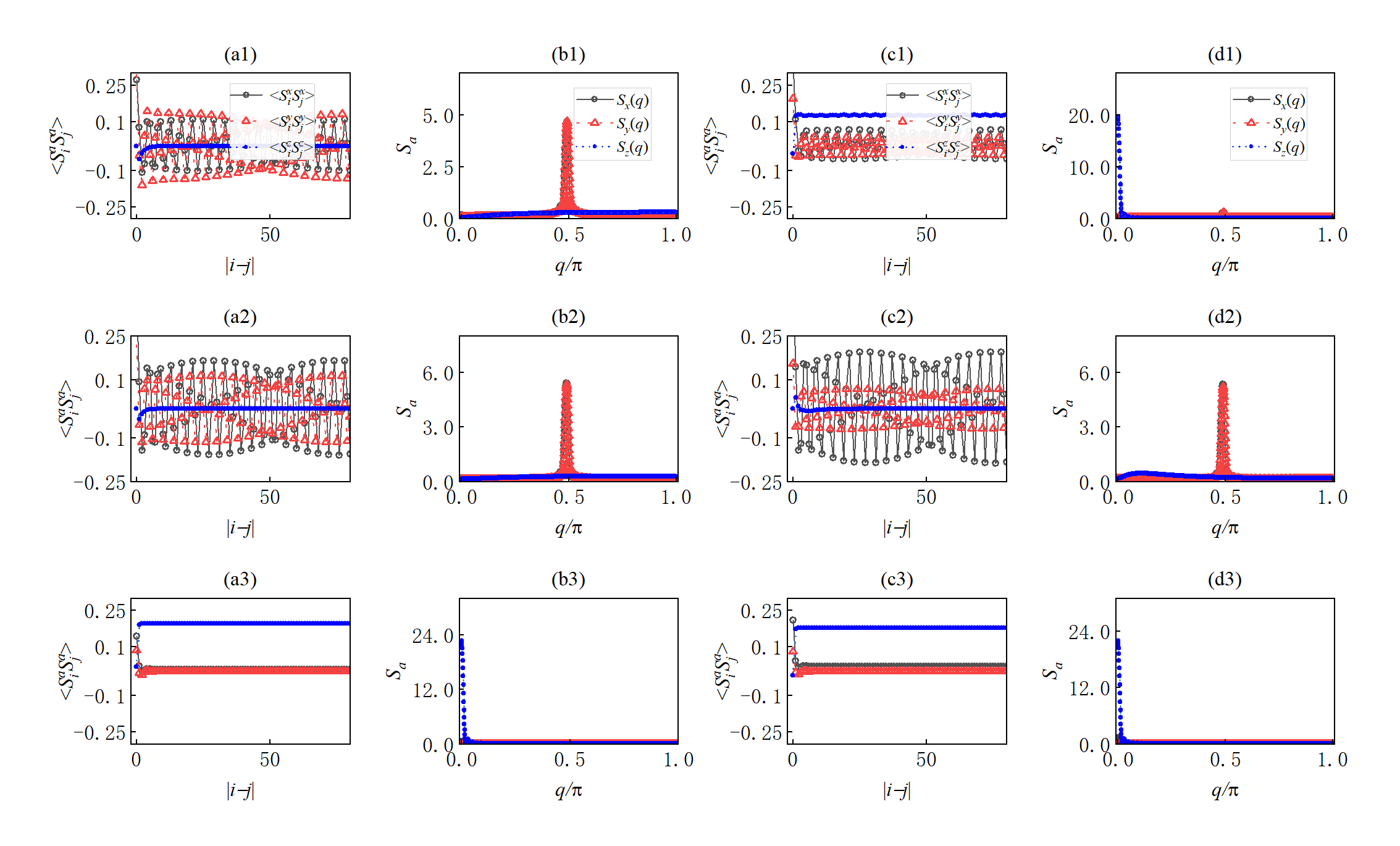}
	\caption{Correlation functions and structure factors for species $A$ and $B$. The three rows correspond to representative parameters (i)-(iii), respectively, signifying the features of the three phases: $A$($xy$-SP)$B$($z$-FM), $A$($xy$-SP)$B$($xy$-SP) and $A$($z$-FM)$B$($z$-FM). (a1-a3) Correlation functions $\left\langle {\hat S_i^a\hat S_j^a} \right\rangle $ ($a=x,y,z$) versus $|i-j|$ for species $A$. The corresponding structure factors ${S^a}$ as a function of wave vector $q$ are plotted in (b1-b3). (c1-c3) Correlation functions for species $B$. Their structure factors are shown in (d1-d3). 	}
	\label{fig4}
\end{figure*}

The phase diagram spanned by $\lambda_A$ and $J_{AB}$ is shown in Fig. 2. Phase boundaries are determined by the variation of order parameters as well as spin-spin correlations, which will be demonstrated in the following. Each region is labeled with the corresponding phases of individual species. For species $A$, there are spiral phase in the $x$-$y$ plane ($xy$-SP) and ferromagnetic phase along $z$ ($z$-FM), as a result of the competition between the DM interaction and the XXZ interaction. With the increase of $U_{AB}^\text{s-e}$ from zero, the phase boundary between $A$($xy$-SP) and $A$($z$-FM) shifts from $\lambda_A=1$. Interestingly, as $U_{AB}^\text{s-e}$ increases along the vertical dashed line, we observe that species $A$ remains in the $xy$-SP, whereas species $B$ experiences a phase transition from the $z$-FM phase to the $xy$-SP phase. Such a phase transition is demonstrated in Fig. 3(a), where long-range correlations are plotted with respect to the inter-chain coupling $J_{AB}$ (with $\lambda_A=0.9$). A clear  phase transition from B($z$-FM) to B($xy$-SP) is identified by the vanishing of $M_z^B$. At this point, $C_z^B$ arises and shows discontinuity in the second-order derivatives. With a strong interaction $U_{AB}^\text{s-e}$, the DM interaction and the inter-chain coupling play dominant roles, leading to $xy$-SP phase for both species. For the regime where species $A$ is in the $z$-FM phase, the two chains are decoupled and the $z$-FM phase is also energetically favored for species $B$ with $\lambda_B>1$. Fig. 3(b) shows the variation of long-range correlations along the horizontal dashed line in Fig. 2. We observe similar behavior of $M_z^A$ and $M_z^B$ arising from zero, representing a simultaneous phase transition from $xy$-SP to $z$-FM for both species. 

To further characterize these phases, we calculate correlation functions and structure factors for ground states with parameters labeled by (i)-(iii) in Fig. 2. Results are demonstrated in Figs. 4, where the first two columns correspond to species $A$ and the last two columns are for species $B$. As shown in Fig. 4(a1, a2), $\langle \hat S_i^{x,A}\hat S_j^{x,A}\rangle$ and $\langle \hat S_i^{y,A}\hat S_j^{y,A}\rangle$ display long-range oscillations, indicating the $xy$-SP phase for species $A$ at representative points (i) $(\lambda_A,J_{AB})=(0.9,0.1)$ and (ii) $(\lambda_A,J_{AB})=(0.9,0.2)$. Their characteristic wavevectors $Q_x$ and $Q_y$ are revealed by the local peaks of $ S_x^A$ and $S_y^A$ in Fig. 4(b1, b2). As shown in Fig. 4(c1, d1), with parameters labeled by (i), species $B$ exhibits a clear long range ferromagnetic order along $z$, accompanied by oscillations of $\langle \hat S_i^{x,B}\hat S_j^{x,B}\rangle$ and $\langle \hat S_i^{y,B}\hat S_j^{y,B}\rangle$, as confirmed by the peak of $S_z^B$ at $Q_z=0$ and a small peak of $S_{x(y)}$ at $q=\pi/2$ with $Q_x=Q_y\simeq 2\phi$. Intuitively, by regarding the spin configuration of species $A$ as a mean-field effect, the inter-chain coupling serves as a spiral field. Species $B$ has additional tendency of orienting the spins following the spiral texture of the $A$-chain. This gives rise to ferromagnetism along $z$ with spiral configuration of spins in the $x$-$y$ plane. In this view, the transition to $xy$-SP for species $B$ resembles the ferromagnetic to paramagnetic phase transition in a transverse field Ising model, which corresponds to the $\mathbb{Z}_2$ symmetry breaking. Therefore we refer to the phase featured by Fig. 4(c1, d1) as the $z$-FM phase. 

With parameters labeled by (ii), the $xy$-SP phase is induced in species $B$, as verified by the similar features of correlation functions in Fig. 4(a2) and (c2). In general, $Q_x=Q_y=2\phi$ is expected in the $xy$-SP phase, indicating a spatial period of $2\pi/\phi$. This can be seen by performing a unitary transformation ($\hat U\hat H_\text{eff}\hat U^\dag$ with $\hat U=\prod_je^{-2ij\phi S_j^{z,A}}$ for species $A$, upon which the DM interaction is converted into an external spiral field 
\begin{equation}
	\hat H_{e}=\sum_j(h_x\hat S_j^{x,A}+h_y\hat S_j^{y,A})
\end{equation}
with $h_x\propto \cos(2\phi j)$ and $h_y\propto \sin(2\phi j)$. When this term is dominant, the coherent spin configuration shares the same period with the spiral field. This is indeed the case for an isolated spin chain \cite{zhang2019magnetic, sun2021effective}. However, in the cases of (i) $A$($xy$-SP)$B$($z$-FM) and (ii) $A$($xy$-SP)$B$($xy$-SP), respectively, we observe that $Q_{x(y)}\simeq 0.491\pi$ and $Q_{x(y)}\simeq 0.489\pi$ for both species. The slight deviation from $2\phi$ may be caused by the interplay between the two spin chains. The last row in Fig. 4  shows typical features of the phase $A$($z$-FM)$B$($z$-FM). In this case, each species is governed by the $z$-component in spin-spin coupling and the two chains are fully decoupled. We mention that with the increase of $J_{AB}$, the boundary between $A$($xy$-SP)$B$($xy$-SP) and $A$($z$-FM)$B$($z$-FM) bends towards larger $\lambda_A$, implying that the inter-chain coupling may enhance the formation of spiral ordering.

\begin{figure}[tbp]
	\centering
	\includegraphics[width=.49\textwidth]{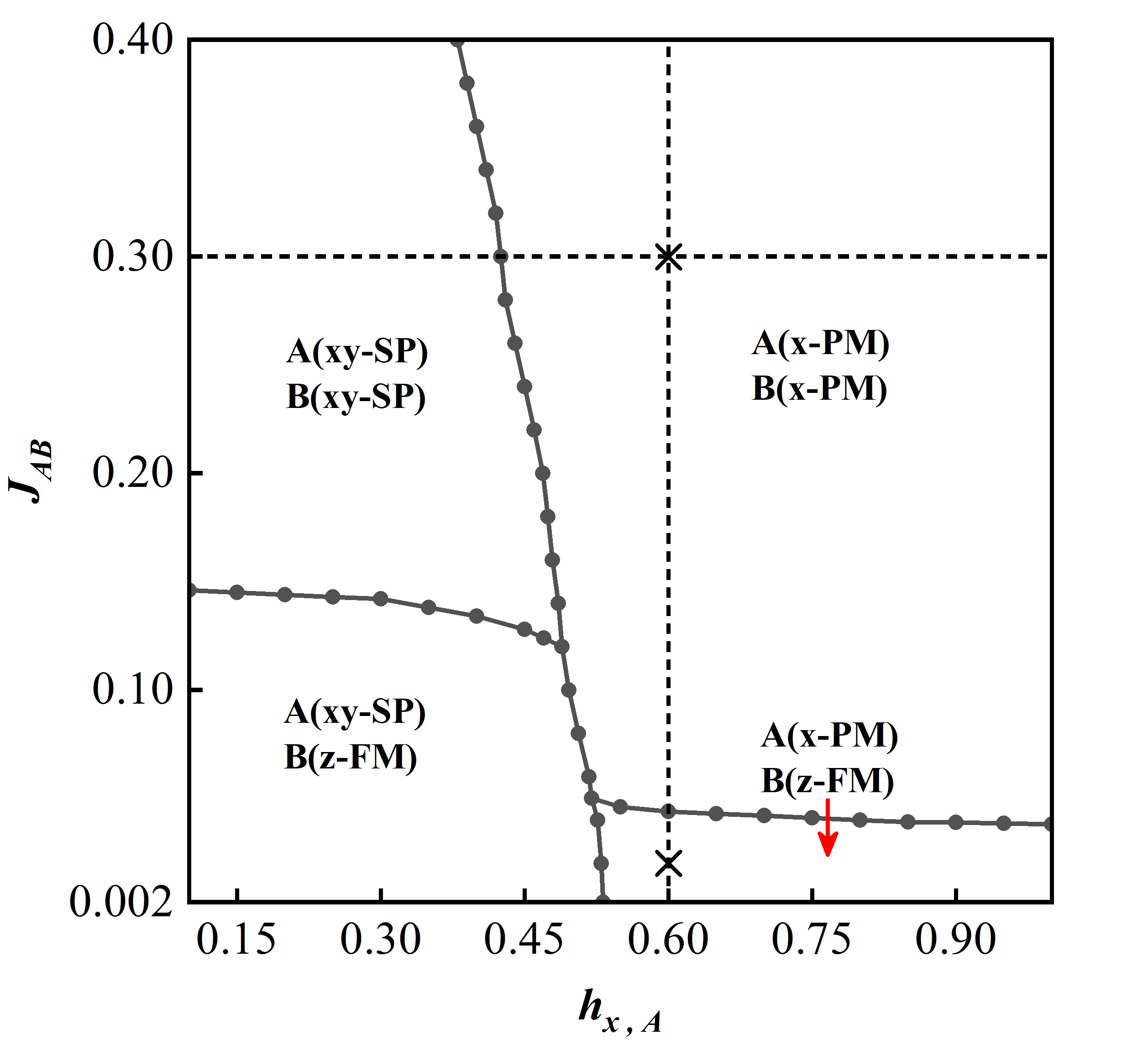}
	\caption{Phase diagram of the spin model with dimensionless parameters $(\phi, \lambda_A, \lambda_B, \beta) = ( 0.25\pi, 0.9, 1.2,0.1)$. The corresponding phases associated with each species are labeled in the diagram. Phase transitions along the dashed lines are demonstrated in Fig. 6. The crosses label typical parameter points which are analyzed in Fig. 7.
	}
	\label{fig5}
\end{figure}

\begin{figure}[tbp]
	\includegraphics[width=0.5\textwidth]{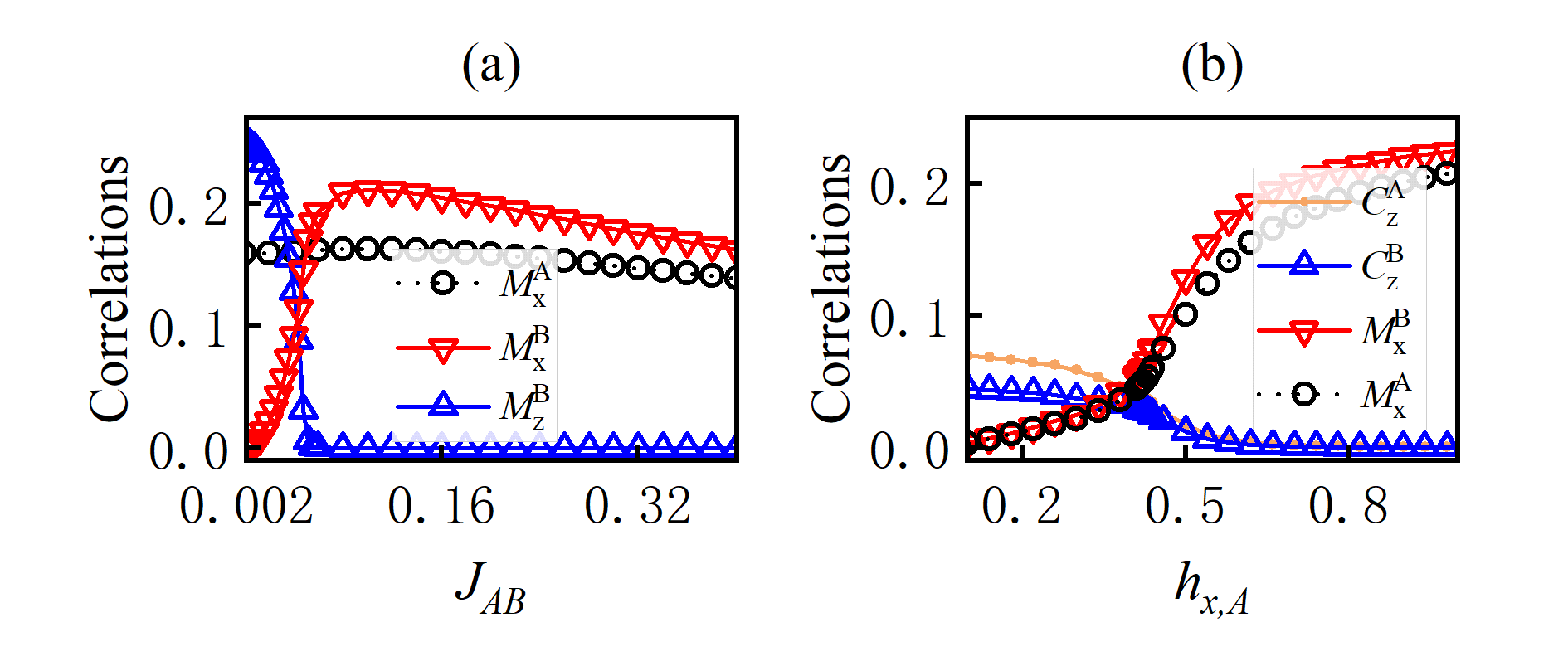}
	\caption{Long-range correlations as a function of $J_{AB}$ and $h_{x,A}$ along the dashed lines in fig. 5. (a) $M_x^A$, $M_x^B$ and $M_z^B$ versus $J_{AB}$ with $h_{x,A} =0.6$. (b)  $C_z^{A(B)}$ and $M_x^{A(B)}$ versus $h_{x,A}$ with $J_{AB}=0.3$. 
	}
	\label{fig.6}
\end{figure}

\begin{figure*}[tbp]
	\includegraphics[width=0.98\textwidth]{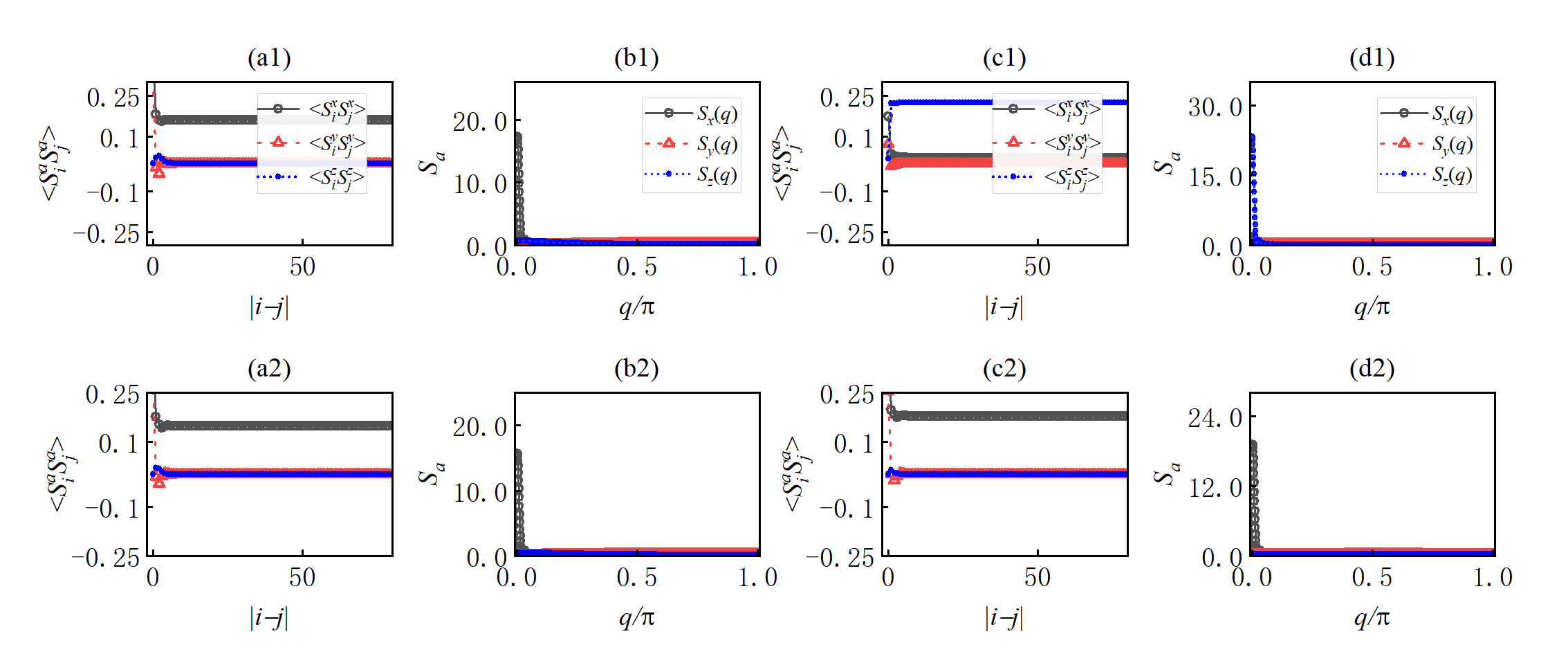}
	\caption{Correlation functions and structure factors for species $A$ and $B$. The two rows correspond to $A$($x$-PM)$B$($z$-FM) and $A$($x$-PM)$B$($x$-PM), respectively. (a1, a2) Correlation functions $\left\langle {\hat S_i^a\hat S_j^a} \right\rangle $ ($a=x,y,z$) for species $A$. The corresponding structure factors ${S_a}$ are plotted in (b1, b2). (c1, c2) Correlation functions for species $B$. Their structure factors are shown in (d1, d2).
	}
	\label{fig.7}
\end{figure*}

\subsection{paramagnetic phase induced by inter-species spin-exchange interaction and transverse field}
 We further study the magnetic phases as a result of the interplay of the transverse field acting on the $A$-chain and the inter-chain coupling. The phase diagram in the $h_{x,A}$-$J_{AB}$ plane is shown in Fig. 5, where we have fixed $\lambda_A=0.9$ and set other parameters the same as in the previous case. For $h_{x,A}\lesssim 0.52$, the phase transition between $B$($z$-FM) and $B$($xy$-SP) still exists, with the boundary slightly bent towards smaller value of $J_{AB}$. The increase of transverse field can eventually drive species $A$ into a paramagnetic phase along $\hat x$ ($x$-PM). As the inter-chain coupling ($J_{AB}$) grows, smaller value of $h_{x,A}$ is required to induce the transition to the $x$-PM phase. For $J_{AB}\gtrsim 0.12$, the two species always undergo a simultaneous $xy$-SP-$x$-PM phase transition. As demonstrated in Fig. 6(a), long-range correlations display second-order discontinuity at critical point. $M_x^A$ ($C_z^A$) and $M_x^B$ ($C_z^B$) exhibit synchronized variation with respect to $h_{x,A}$ (Along the dashed line in Fig. 5 with $J_{AB}=0.3$). When species $A$ is in the $x$-PM phase, the competition between the $S^{z,B}$-$S^{z,B}$ spin exchange interaction and the $J_{AB}$-term leads to a phase transition from $B$($z$-FM) to $B$($x$-PM). This can be seen from Fig. 6(b), where $M_z^B$ vanishes at critical point with finite $M_x^A$. Again, by regarding the polarized spins in species $A$ as a mean field effect, the coupling $S_j^{x,A}S_j^{x,B}$ in the $J_{AB}$-term provides an effective transverse field. In this sense, the ferromagnetic to paramagnetic phase transition in species $B$ is in the universality class of a transverse-field Ising model. 

We present correlation functions and structure factors for both species in Fig. 7. At representative point in the $A$($x$-PM)$B$($z$-FM) phase, species $A$ is characterized by long-range correlation in $\langle \hat S_i^x \hat S_j^x\rangle$ and a local peak of $S_x$ at $Q_x=0$, as shown in Fig. 7(a1, b1). Species $B$ is featured by nearly saturated ferromagnetic correlation in $\hat z$ direction, as confirmed by the sharp peak of $S_z$ [see Fig. 7(c1) and (d1)]. Here a finite ferromagnetic correlation along $\hat x$ exists due to the presence of the effective transverse field generated by the inter-chain coupling. In the $A$($x$-PM)$B$($x$-PM) phase, as demonstrated in Fig. 7(a2-d2), both species are characterized by long-range correlation in $\langle S_i^x S_j^x\rangle$ with a peak of $S_x$ at $Q_x=0$.

\section{Summary}\label{IV}
The dual-species spinor lattice system provides a versatile and rich platform for quantum simulation, which enables exploring exotic spin Hamiltonians and engineering non-trivial magnetic states through inter-species coupling induced physics. We have investigated the Mott insulating phases of dual-species bosonic spinor lattice gases. A synthetic flux is applied to one species by implementing spin-dependent tunneling phases. In the presence of inter-species spin exchange interactions, the system manifests a two-layer ladder in the spirit of synthetic dimension. The synthetic flux represents an effective SCO and gives rise to the DM interaction in the deep Mott regime. We obtain an unconventional spin model composed of two spin chains coupled by spin exchange interaction in $\hat x$ and $\hat y$ directions. This system, tuned by varying intra- and inter-species interactions, tunneling strengths and synthetic flux, reveals rich magnetic phases with emergent properties not found in single-component systems. By employing the TEBD method based on TN representations, we obtain typical phase diagrams with synchronous and asynchronous phase transitions for the two species. We find that the interplay of the DM interaction in one species and the inter-species spin exchange interaction (\textit{i.e.}, the inter-chain coupling) can induce spiral ordering in the other species. Analogously, the combined action of a sufficient transverse field and the inter-species spin exchange interaction leads to paramagnetic phase for both species. These effects are consistent with the presence of induced SOC by spin exchange interaction in a superfluid mixture. 

Our work establishes a comprehensive framework for understanding magnetic phases induced by inter-chain coupling. A natural extension is to explore higher dimensional models. For example, a 2D dual-species lattice system can be mapped to a two-layer spin model in the deep Mott regime, which may give rise to exotic magnetic orders such as the (induced) vortex lattice or the skyrmion. Despite their increased complexity, multi-component systems host phases beyond the scope of single-component systems. 

\section{Acknowledgements}
This work is supported by NSFC (Grant No. 12304179)
, the China Postdoctoral Science 
Foundation (Grant No. 2019M650025) and Sichuan Normal University College Students' Innovation and Entrepreneurship Training Program(Grant No. 202510636005).

\bibliography{reference1.bib}

\end{document}